\documentstyle[prl,aps,epsf,twocolumn,floats]{revtex}
\begin{document} 
\title{Detecting and analysing nonstationarity in a time series with nonlinear
cross--predictions}

\author{Thomas Schreiber\\
      Physics Department, University of Wuppertal, D--42097 Wuppertal,
      Germany}
\date{Phys.\ Rev.\ Lett.\ {\bf 78}, 843 (1997)}
\maketitle
\begin{abstract}
We propose an informal test for stationarity in a time series which checks for
the compatibility of nonlinear approximations to the dynamics made in different
segments of the sequence. The segments are compared directly, rather than via
statistical parameters. The approach provides detailed information about
episodes with similar dynamics during the measurement period. Thus physically
relevant changes in the dynamics can be followed.
\end{abstract}

Almost all methods of time series analysis, traditional linear or nonlinear,
must assume some kind of stationarity. Therefore, changes in the dynamics
during the measurement period usually constitute an undesired complication of
the analysis. There are however situations where such changes represent the
most interesting structure in the recording. For example,
electroencephalographic (EEG) recordings are often taken with the main purpose
of identifying changes in the dynamical state of the brain. Such changes occur
e.g. between different sleep stages, or between epileptic seizures and normal
brain activity. In this letter we propose an approach to the
study of potentially nonstationary signals which does not only provide a
powerful test for stationarity but also allows for a time resolved study of the
dynamical changes. While testing for stationarity might appear to be a
technical problem of time series analysis, the analysis and understanding of 
nonstationary signals is a topic of current research in many areas of science.

A number of statistical tests for stationarity in a time series have been
proposed in the literature.  Most of the tests we are aware of are based on
ideas similar to the following: Estimate a certain parameter using different
parts of the sequence. If the observed variations are found to be significant,
that is, outside the expected statistical fluctuations, the time series is
regarded as nonstationary. In many applications of linear (frequency based)
time series analysis, stationarity has to be valid only up to the second
moments (``weak stationarity''). Then, the obvious approach is to test for
changes in second order quantities, like the mean, the variance, or the power
spectrum. See e.g.~\cite{priestley} and references therein. Nonlinear
statistics which can be used include higher order correlations, dimensions,
Lyapunov exponents, and binned probability distributions.\cite{kurths}

Stationarity can also be tested for without comparing running statistical
parameters. One such test which is particularly useful in the context of
correlation dimension estimates is the space--time separation plot introduced
in Ref.~\cite{provenzale}. Also the recurrence plot of Ref.~\cite{recurr} and
the method proposed in~\cite{kennel} provide related information. However,
these algorithms do not allow for a time resolved study. Other material
concerning nonstationarity in a nonlinear setup is found in Ref.~\cite{B1}.

In the following, a novel approach is taken which is based on the similarity
between parts of the time series themselves, rather than the similarity of {\em
parameters} derived from the time series by local averages.  In particular, the
(nonlinear) cross--prediction error, that is, the predictability of one segment
using another segment as a data base, will be evaluated. This concept is
particularly useful if nonstationarity is given by changes of the shape of an
attractor while dynamical invariants remain effectively unchanged. Other
statistics which measure the similarity of time series can be used
alternatively.

Let $\{x_n,\quad n=1,\ldots,N\}$ be a time series which is split the series
into adjacent segments of length $l$, the $i$--th segment being called
$S_i^l=\{x_{(i-1)l+1},\ldots,x_{il}\}$.  Traditionally, a statistic $\gamma_i$
is now computed for each such segment. It is then tested if the sequence
$\{\gamma_i\}$ is constant up to statistical fluctuations. How this is done
depends on what we know about the properties of the statistic $\gamma$, in
particular its probability distribution.  Alternatively, one can compare
statistics computed on segments to values obtained from the full sequence. Note
that $\gamma$ is typically a scalar but vectors like binned distributions can
also be used. In this paper, we will take a different approach and use
statistics defined on pairs of segments, $\gamma_{ij}=\gamma(S_i^l,S_j^l)$, in
particular the cross--prediction error.

Statistical testing with nonlinear parameters $\gamma$ is difficult because we
can assume very little about the statistical properties of $\gamma$. Estimators
of dimensions and Lyapunov exponents do not usually follow normal
distributions. Mean prediction errors are composed of many individual errors
and thus more likely to be normal. However, the empirical errors are not
expected to be independent which complicates the estimation of the variance of
$\gamma$. By using statistics $\gamma_{ij}$ on pairs of segments we increase
the number of parameters computed at a fixed number of segments from $N/l$ to
$(N/l)^2$. It can be argued that we gain largely redundant information for the
purpose of statistical testing since the $\gamma_{ij}$ for different $i,j$ are
not expected to be independent. However, we will be able to detect different
and more hidden kinds of nonstationarity. We can get a more detailed
picture about the nature of the changes and, in particular, locate segments
of a nonstationary sequence which are similar enough for the purpose of our
analysis and which can therefore be analysed together.

In principle, $\gamma(S_i,S_j)$ can be any quantity which is sensitive to
differences in the dynamics in $S_i$ resp.~$S_j$. Examples of such quantities
can be found in Refs.~\cite{holger} and~\cite{lou}. For the application we have
in mind, theoretical rigor in the definition of $\gamma$ is less important than
robustness and the possibility to obtain a stable estimate on rather short
segments $S_i,S_j$. One statistic which meets these criteria is the error of a
simple nonlinear prediction algorithm. Predictions with locally constant
approximations yield stable results for sequences of a few hundred points or
less. Global nonlinear predictions can be performed with even less points,
provided the global ansatz is chosen properly. Here we want to avoid the latter
nontrivial requirement. More attractive from the theoretical point of view is
the crosscorrelation integral defined in Ref.~\cite{holger}. However, it
requires longer segments and a low noise level in order to obtain stable
results without manual evaluation of scaling plots. 

Let us again stress that the main point of this letter is to exploit the
information contained in the relative statistics $\gamma(S_i,S_j)$,
in addition to that contained in the diagonal terms $\gamma(S_i,S_i)$.
Many nonlinear statistics can be naturally generalized to relative quantities.
We mentioned cross--prediction errors and the crosscorrelation integral.
Lyapunov exponents might be generalized by measuring the divergence of pairs
of trajectories, one taken from $S_i$, one from $S_j$.

Let us define the {\em cross--prediction error} $\gamma_{ij}$ we will use as a
statistic to compare segments. It is computed as follows. Let
$X\equiv\{x_n,\quad n=1,N_X\}$ and $Y\equiv\{y_n,\quad n=1,N_Y\}$ be two time
series and $m$ be a small integer denoting an embedding dimension.  From both
time series we can form embedding vectors $\{\vec{x}_n,\quad n=m,N_X-1\}$
resp. $\{\vec{y}_n,\quad n=m,N_Y-1\}$ in the same $m$ dimensional phase space,
where $\vec{x}_n=(x_{n-m+1},\ldots,x_n)$. Further let us fix a length scale
$\epsilon$. For each $\vec{y}_n$ we want to make a prediction one step into the
future, that is, given $\vec{y}_n=(y_{n-m+1},\ldots,y_n)$ we want to estimate
$y_{n+1}$, using however $X$ as a data base. A locally constant approximation
to the dynamics relating $\vec{x}_n$ and $x_{n+1}$ yields the estimate
$$ \hat y_{n+1}^X =
   \frac{1}{|{\cal U}_{\epsilon}^X(\vec y_n)|} \sum_{\vec x_{n'}\in {\cal
   U}_{\epsilon}^X(\vec y_n)} x_{n'+1}
\,.$$
In the above formula, ${\cal U}_{\epsilon}^X(\vec y_n)=\{\vec x_{n'}:
\|\vec x_{n'}-\vec y_n\|<\epsilon\}$ is an $\epsilon$--neighborhood of 
$\vec y_n$, formed however within the set $X$. $|{\cal U}_{\epsilon}^X(\vec
y_n)|$ denotes the number of elements in that neighborhood. For isolated points
with empty neighborhoods we take the sample mean of the segment $X$ as an
estimate $\hat y_{n+1}^X$. This or similar schemes are widely used for
prediction and noise reduction.  Our formulation however leaves room for the
possibility that the approximation to the dynamics is performed on a different
data set $X$ than the actual prediction. If we take $X$ and $Y$ to be the same
but exclude from ${\cal U}_{\epsilon}^Y(\vec y_n)$ all $2m-1$ vectors which
share components with $\vec y_n$, then $\hat y_{n+1}^Y$ is an ordinary
out--of--sample prediction of $y_{n+1}$.\cite{foot:out-of-sample} The root mean
squared prediction error $\gamma(X,Y)$ of the sequence $Y$, given $X$, is
defined by
$$   \gamma(X,Y)=\sqrt{{1\over N_Y-m}\sum_{n=m}^{N_Y-1} 
      (\hat y_{n+1}^X-y_{n+1})^2}\,.$$
For $X=Y$, this is the usual take--one--out, out--of--sample prediction error.
$\gamma(X,Y)$ probes in how far the locally constant approximation to the
dynamics of $X$ is suitable to predict values in $Y$. For a stationary time
series, we expect that $\gamma(S_i^l,S_j^l)$ is independent of $i$ and $j$
unless the coherence time of the process is longer than~$l$. If there is
variability in the sequence on time scales longer than~$l$, be it due to a slow
variable or due to a changing parameter, the diagonal terms
$\gamma(S_i^l,S_i^l)$ will be typically smaller than those with $i\neq j$. 

Note that in general $\gamma(X,Y)\neq\gamma(Y,X)$. In particular, if the
attractor of $Y$ is embedded within the attractor of $X$, for example if $Y$
forms a periodic orbit which is present as an unstable orbit in $X$, points in
$Y$ can be well predicted using $X$ as a data base. However, $Y$ does not
contain enough information to predict all points in $X$. While the asymmetry of
$\gamma_{ij}$ can provide valuable insights, it may also be confusing in some
cases.  One can then use a symmetrized statistic like 
$\gamma_{ij}+\gamma_{ji}$.

\begin{figure}
\centerline{\epsffile{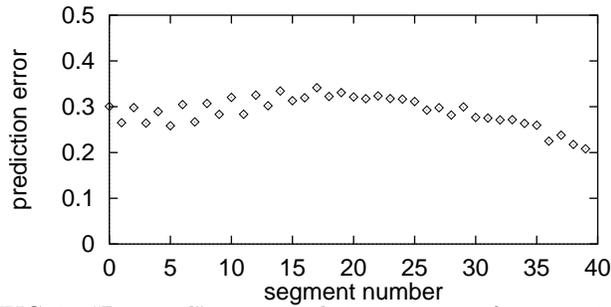}}
\caption{``Diagonal'' cross--prediction error $\gamma_{ii}$ for a
nonstationary sequence of the generalized baker map with $\alpha=0.4$ and
$\beta=n/N$. The total signal of $N=40000$ is split into 40 segments of length
1000.\label{fig:diagonal}}
\end{figure}

Let us illustrate the method with a numerical example, a generalization of the
well known ``baker map''
$$  \begin{array}{ll} {\rm if}\; v_n\leq\alpha: &
                \begin{array}{rcl} u_{n+1}&=&\beta u_n \\
                                   v_{n+1}&=&v_n/\alpha \end{array}\\
                      {\rm if}\; v_n>\alpha: &
               \begin{array}{rcl} u_{n+1}&=&0.5+\beta u_n \\
                              v_{n+1}&=&(v_n-\alpha)/(1-\alpha)\,,\end{array}\\
            \end{array}$$
defined for $v_n\in [0,1],\quad \alpha,\beta\in ]0,1[$. For this
piecewise linear mapping, the two Lyapunov exponents can be computed
analytically (see e.g.~\cite{schuster}):
\begin{eqnarray*}
   \lambda_1&=&\alpha \log{1\over\alpha}+(1-\alpha)\log{1\over 1-\alpha}\\
   \lambda_2&=&\log\beta\,.
\end{eqnarray*}
By varying $\beta$ only, we can create sequences with different dynamics but
with the same maximal Lyapunov exponent. Indeed, we will generate a
nonstationary time series by varying $\beta$ slowly with time: $\beta=n/N$. We
keep $\alpha=0.4$ fixed and measure $N=40000$ points by recording $u+v$. From
this we subtract the running mean and we normalize to unit running
variance. The actual time series is thus:
$$ x_n={w_n-\langle w\rangle_k \over 
      \sqrt{\langle (w-\langle w\rangle_k)^2\rangle_k}},\qquad w_n=u_n+v_n\,,$$
where $\langle\cdot\rangle_k$ denotes the average over indices
$n'=n-k,\ldots,n+k$. Here $k=50$. The nonstationarity in the sequence is very
hard to detect since many observables remain unchanged. The running mean and
variance are constant up to finite sample fluctuations, autocorrelations show
only very small variation, finite time estimates of the largest Lyapunov
exponent essentially do not change.  Figure~\ref{fig:diagonal} shows the
nonlinear prediction error $\gamma_{ii}$ for 40 segments of length 1000
each. We used an $m=2$ dimensional embedding and neighbourhoods of radius
$\epsilon=0.25$. Only towards the end of the sequence one could suspect that
something is changing.

\begin{figure}
\centerline{\epsffile{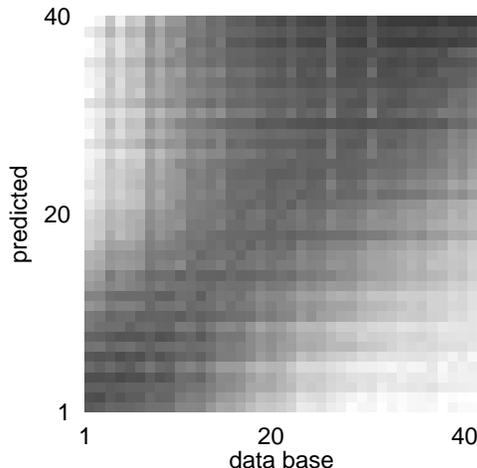}}
\caption{Mutual predictions between sections of length 1000 for the baker map
time series used in Fig.~\ref{fig:diagonal}.\label{fig:ls}}
\end{figure}

The parameter drift is however revealed by Cross--predictions using one segment
$S_i^l$ of length $l=1000$ as a data base to predict values within another
segment $S_j^l$, as can be seen in Fig.~\ref{fig:ls}. Prediction errors are
encoded as gray scales. Black is used for $\gamma_{ij}\leq 0.3$, white for
$\gamma_{ij}\geq 0.8$, and linear shading in between. Predictability degrades
rapidly with the temporal distance of the segments.

As a realistic example, we study a recording of the breath rate of a human
patient during almost a whole night (about 5~h), measured twice a second. The
data is part of data set B from the Santa Fe Institute time series contest held
in 1992. It is described in Ref.~\cite{gold}. Obviously, conditions cannot be
assumed to be constant during a night's sleep. Changes of the calibration and
the instantaneous variance, as well as the linear autocorrelations are easy to
detect by standard methods. In order to emphasize that the algorithm is
sensitive to changes in the nonlinear structure, we subtract from the data the
running mean and divide it by the running root mean squared amplitude. Further,
all prediction errors are normalized to the error of the best linear AR(1) 
model.

\begin{figure}
\centerline{\epsffile{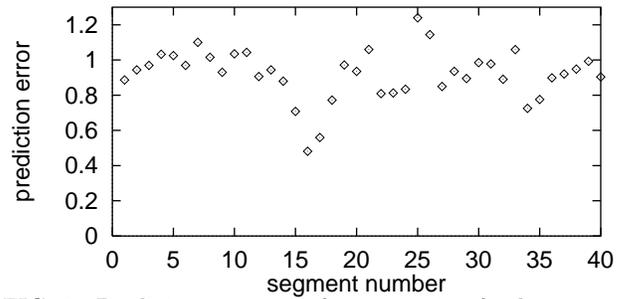}}
\caption[]{Prediction error $\gamma_{ii}$ for segments
   of a long, nonstationary recording of the breath rate of a human.\cite{gold}
   The set was split into 40 segments of 850 points (425~secs.)
   each. Errors are normalized to the error of the best AR(1) model. 
   Considerable fluctuations are present but there is no indication of a
   qualitative difference between the first and the second half of the
   recording.\label{fig:B2diagonal}}
\end{figure}

\begin{figure}
   \centerline{\epsffile{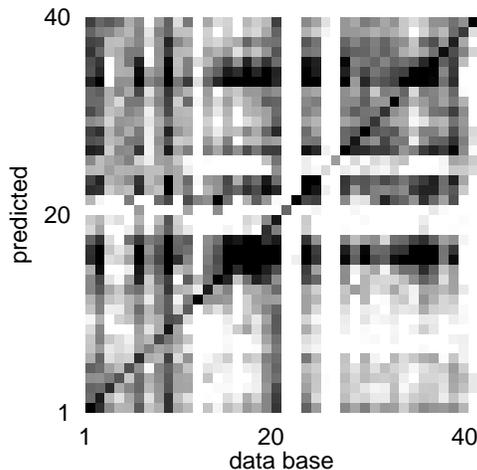}} 
   \caption[]{Cross--prediction errors for segments
   of a recording of the breath rate of a
   human. The figure shows that there is a qualitative change in the dynamics
   around segment 15.\label{fig:b}} 
\end{figure}

In order to detect nontrivial changes, we split the recording into 40
nonoverlapping segments $S_i$ of 850 points (425~sec.)
each. Cross--predictions are performed using $m=2$, and $\epsilon$ is chosen to
be $0.25$ (at unit rms amplitude).  In Fig.~\ref{fig:B2diagonal} we show the
(auto--) prediction error $\gamma_{ii}$ as a function of segment number. There
are some fluctuations, most prominent are the lower errors for segments 15--18.
The cross--prediction error is shown in Fig.~\ref{fig:b} as grey shades. Black
means $\gamma\leq 0.9$, white $\gamma>1.3$, linear grey shades are used in
between. Except from the lower errors for $i=j$ (see
note~\cite{foot:out-of-sample}), we see that there is a transition around one
third of the recording: segments up to about 15 are less useful for predictions
of segments after about 20 and vice versa. That segments 15--18 are different
was aparent from Fig.~\ref{fig:B2diagonal} already. For this data set, most
nonlinear tests are able to detect that nonstationarity is present. The main
advantage of the present method is that it provides more detailed, time
resolved information than just a statement that nonstationarity has been found.

The algorithm as described here contains a few parameters which have to be
chosen appropriately for each data set. The embedding dimension $m$ and
neighborhood size $\epsilon$ should yield good overall predictions.  The
segment size is determined by the tradeoff between statistical stability of
$\gamma_{ij}$ for long segments and finer time resolution for short segments.
A slight advantage may be gained by the use of overlapping segments.  Other
relative statistics than cross--predictions may be used and the table of the
$\gamma_{ij}$ may be interpreted by other means than grey scale plotting.  In
particular, ongoing research is devoted to the evaluation of $\gamma_{ij}$
in terms of cluster analysis.

In a nonlinear setting, for instance if it is planned to apply algorithms from
the theory of deterministic chaos to a time series, weak stationarity (constant
second moments) is certainly not enough.  Let us further remark that the
widespread notion that the {\em system} which produces the time series must
remain unchanged during the time of measurement is neither a necessary nor a
sufficient condition for stationarity.  The reason is that there is no {\em a
posteriori} distinction between a system parameter (to remain constant) and a
variable (which may evolve in time). Thus a system with a rapidly fluctuating
parameter may yield a stationary time series because these fluctuations can be
averaged over, while a system with constant parameters can produce signals
which for all practical work must be considered nonstationary. An example for
the latter case is given by intermittency where the time evolution of some
variables may become arbitrarily slow. For processes, stationarity can be
defined by requiring that the joint probability distribution remains
constant. Given a finite time series, this probability distribution can only be
estimated up to statistical fluctuations.  It is problematic to define
stationarity on the base of such an estimate.  In this paper we have taken a
rather pragmatic point of view and call a signal stationary if anything which
changes in time (no matter if we call it a variable or a parameter) does so on
a time scale such that the changes average out over times much smaller than the
duration of the measurement.

We were able to detect changes in the dynamics of a system even if scalar
statistics do not change significantly. The proposed method is meant to augment
known tests for stationarity, in particular since it includes the possibility
to find interrelations and similarities between different parts of a time
series.

We thank Holger Kantz and Peter Grassberger for useful comments.  This work was
supported by the SFB 237 of the Deutsche Forschungsgemeinschaft.


\begin{thebibliography}{10}
\bibitem{foot:out-of-sample} Since consecutive individual prediction errors
   are not independent, temporal neighbors of $\vec y_n$ closer than a typical
   correlation time must be excluded in order to obtain an out--of--sample
   error.
\end{thebibliography}
\end{document}